\documentclass[%
reprint,
nofootinbib,
 amsmath,amssymb,
 aps,
]{revtex4-1}

\usepackage{graphicx}
\usepackage{dcolumn}
\usepackage{bm}
\usepackage{hyperref}
\usepackage[mathlines]{lineno}

\begin{document}

\preprint{APS/123-QED}

\title{Positioning systems in Minkowski  space-time: Bifurcation problem \\ and observational data}

\author{Bartolom\'{e} Coll}
 \email{bartolome.coll@uv.es}
\author{Joan Josep Ferrando}
 \email{joan.ferrando@uv.es}
\author{Juan Antonio Morales-Lladosa}%
 \email{antonio.morales@uv.es}
\affiliation{%
Departament d'Astronomia i Astrof\'{\i}sica, Universitat de
Val\`encia, 46100 Burjassot, Val\`encia, Spain.
}%


\begin{abstract}

In the framework of relativistic positioning systems in Minkowski space-time, the determination of the inertial coordinates of a user  involves the 
{\em bifurcation problem} (which is the indeterminate location of a pair of different events receiving the same emission coordinates). 
To solve it, in addition to the user emission coordinates and the emitter positions in inertial coordinates, it may happen that 
the user needs to know {\em independently} the  orientation of its emission coordinates. Assuming that the user may observe the relative 
positions of the four emitters on its celestial sphere, an observational rule to determine this orientation is presented.  
The bifurcation problem is thus solved by applying this observational rule, and consequently, {\em all} of the parameters in the general expression of the coordinate transformation 
from emission coordinates to inertial ones may be computed from the data received by the user of the relativistic positioning system. 

\end{abstract}


\maketitle



\section{Introduction}
\label{intro}

To locate  the users%
\footnote{The word ``user" here denotes any person or device able to receive the pertinent emitted data from the relativistic positioning system and to extract from it the corresponding information. For short, we shall refer to the user as ``it". \label{user}}
of a Global Navigation Satellite System (GNSS), several geometric methods and 
algebraic algorithms have been developed in the past 
\cite{Schmidt-72, Bancroft-85, Kreuse-87, Abel-Chaffee-91, Chaffee-Abel-94} 
that are still in use \cite{Strang-Borre-97, Juang-Tsai-09}. 
Basically, the algebraic statement of the {\it location problem} is rather simple: to find 
the events where the emission light cones of four broadcast signals intersect. 
Of course, this idea is implicit in the Bancroft algorithm \cite{Bancroft-85} and other similar ones \cite{Kreuse-87}. In fact, Abel and 
Chaffee \cite{Abel-Chaffee-91, Chaffee-Abel-94}
used Minkowskian algebra to state the problem properly, making apparent
that the more Lorentzian a description is, the more clear algorithm is performed.

However, in a full relativistic framework (cf. \cite{Bahder-2001, Coll-ERE-2000, Coll-Bucarest-2002, Rovelli-2002, Hehl-2002, Bahder-2003,D4a}), 
and even in the case of the flat space-time, an explicit form
of the solution of the location problem for arbitrary emitters has not been obtained
until recently \cite{emission-1}.

In \cite{emission-1}, an exact relativistic formula giving the inertial coordinates of an event in terms of the 
received emission coordinates is obtained. This formula applies 
in {\em all} the emission coordinate region and involves the orientation of the emission coordinates of the user. 
Nevertheless, there exists an inherent limitation 
on the applicability of this formula: only the users in a certain region (named the central region, see Sec. \ref{subsec-4b}) of a positioning
system can obtain the orientation from the sole {\it standard emission data}, that is to say, from the sole set of the 
positions of the four emitters in inertial coordinates and of the emission coordinates of the user. 
Consequently, only these restricted users are able to locate themselves in inertial coordinates.

Here, assuming that the users out of the central region may observe the relative positions of the 
four emitters on their celestial sphere, we will give a simple rule allowing {\em any} user of the positioning system 
to locate itself in inertial coordinates. To show that, we will see that the orientation of the emission coordinates of a user is related to the relative 
positions of the emitters of the positioning system on the celestial sphere of the user.

In building current GNSS models, the usual assumption consists in picking out an approximate numerical solution. 
But, because gravitational effects are not taken into consideration at the considered leading order, one should start 
from the best accurate solution that nowadays we know. Such a solution is precisely the simple, exact,  and covariant 
formula, found  in Ref. \cite{emission-1} and improved here,  giving  the location of a user of a relativistic positioning system 
in  Minkowski space-time. 

Let us remark that our result not only concerns GNSS around the Earth, but also general (relativistic) positioning systems anywhere in the 
Solar System or elsewhere. It is true that for most, but not all, of the present applications of the GNSS the users are near the Earth's surface. 
Therefore, they are usually in the central region of the satellites they detect, so that additional data (and in particular our 
observational rule) are not necessary. But for other applications of the GNSS as well as for general positioning systems, 
our observational rule may be a simple way to solve the bifurcation problem and hence, the location one.

  \subsection{Outline of the paper}
  \label{subsec-1a}

The paper is organized as follows. In Sec. \ref{sec-2}, the inertial coordinates of a user are expressed in terms of the
emitter configuration and the orientation of the positioning
system. This provides a covariant formula for the transformation
from emission to inertial coordinates. An analysis of the solution in terms of the configuration of the emitters is also presented. 
In Sec. \ref{sec-3}, some properties of the border between the two emission coordinate domains are obtained, and an 
observational rule to detect it is remembered. Section \ref{sec-4} is devoted to define the genuine regions and coordinate 
domains involved in the problem. We stress the geometrical meaning  
of the coordinate transformation formula in connection with these regions. 
In particular, we show  that in the central region of the positioning system the orientation is computable from the sole 
standard emission data. In Sec. \ref{sec-5} we discuss the bifurcation problem (nonuniqueness of solutions 
in the determination of the location) which is related to the existence of regions 
whose events can  not be located from the sole standard emission data. We give an {\it observational rule} 
to solve the above indetermination problem. This rule
allows us to determine, at any event in the emission region, the orientation of 
the emission coordinates of the user
from the {\em observational data} of the relative positions 
of the emitters on the celestial sphere of any user at this event. 
The concluding Sec. \ref{sec-concluding} is devoted to summarize and discuss  the results. The used notation is explained in an appendix.

Some preliminary results of this work were presented 
at the Spanish Relativity meeting ERE-2010 \cite{NosERE10b}.

 \subsection{Relativistic positioning terminology: Brief compendium}
 \label{subsec-1b}

As pointed out in Ref. \cite{Coll-ERE-2000}, {\it relativistic positioning systems} 
\cite{Bahder-2001, Coll-ERE-2000,  Coll-Bucarest-2002, Rovelli-2002, Hehl-2002, Bahder-2003} 
and the {\it emission coordinates} \cite{D4a, NwReEC} they
realize are essential elements to develop the relativistic theory
of the GNSS. Starting from scratch, we  present here a compendium of  basic definitions about this specific subject. Anyway, we consider 
these definitions necessary not only to make this paper self contained, but also as an 
 incipient piece of concepts to deal with GNSS in a full relativistic perspective.

{\it Relativistic positioning system:}  set of four emitters $A$ ($A=1,2,3,4$), of  worldlines $\gamma_A(\tau^A)$, broadcasting their
respective proper times $\tau^A$ by means of electromagnetic signals.%
\footnote{For simplicity the proper time is taken here, but any other time is valid. For example, the Global Positioning System (GPS) broadcasts 
the GPS time, a time which, roughly speaking, coincides up to a fixed shift with the International Atomic Time (TAI), a sort of mean proper time on 
the Earth surface.\label{GPS-time}}%

{\it Emission coordinates of an event:} the four times $\{\tau^A\}$ which are received at each event reached by the emitted signals.%
\footnote{Emission coordinates have received different appellations in the past 
(see  \cite{NosERE10a} for a brief and critical account). \label{emico}}

{\it Configuration of the emitters for an event x:} set of four events $\{\gamma_A(\tau^A)\}$ of the emitters  
at the emission times  $\{\tau^A\}$ received at $x$.

{\it Emission region:} set $\cal{R}$ of events reached by the four signals broadcast by the  positioning system. Every $ x \in \cal{R}$ is  
labelled with the corresponding emission coordinates $\{\tau^A\}$.

{\it Characteristic emission function:}  map $\Theta$ that to every $ x \in \cal{R}$ associates its emission coordinates, that is
$\Theta(x) = (\tau^A)$.

The characteristic emission function describes the action of a positioning system and, hence, represents it.

{\it Emission coordinate region:} subset ${\cal C}$ of the emission region ${\cal R}$ where the gradients $d \tau^A$ are well defined and
linearly independent. 

The emitter worldlines are excluded from  ${\cal C}$ because every
$d\tau^A$ is not defined at the emission event $\gamma_A(\tau^A)$ (this event being the vertex of the emission light cone $\tau^A =  Constant$).

{\it Orientation of a relativistic positioning system at  the event $x$:} orientation of its emission coordinates at $x$. 
It is given by the sign ${\hat \epsilon}$ of the Jacobian determinant $j_\Theta(x)$ of $\Theta$ at $x$,  
${\hat \epsilon} \equiv \textit{sgn} \, j_\Theta(x)$.

In terms of the gradients of the emission coordinates, one has 
\begin{equation} \label{jacobia}
{\hat \epsilon} = \textit{sgn} [*(d\tau^1 \wedge d\tau^2 \wedge d\tau^3 \wedge
d\tau^4)]
\end{equation}
where  $*$  stands for the Hodge dual operator, and $\wedge$ is the exterior
product (see Appendix  \ref{apendix} for transcription into index notation).


\section{The location problem in Minkowski space-time}
\label{sec-2}

Suppose a given specific coordinate system $\{x^{\alpha}\}$ covering the emission region ${\cal R}$,  let $\gamma_A(\tau^A)$ 
be the worldlines of the emitters referred to this particular coordinate system, and let $\{\tau^A\}$ be the values of the emission coordinates 
received by a user. The data set  $E  \equiv$ $ \{\gamma_A(\tau^A), \{\tau^A\} \}$ is called the {\em standard emission data set}.  

The {\em location problem with respect to} $E$, also called the {\em standard location problem} for short, is the problem of finding 
the coordinates $\{x^{\alpha}\}$  of the user from the sole data $E$. 

In Ref. \cite{emission-1}, the above standard location problem was analyzed for {\em arbitrary} relativistic positioning systems in Minkowski space-time, 
assuming that the specific coordinate system  $\{x^{\alpha}\}$ is an {\em inertial} one. There, 
the explicit expression $x^{\alpha} = \kappa^{\alpha}(\tau^A)$ was found, 
giving the coordinate transformation from emission coordinates to inertial ones (Eq. (\ref{generaltransf}) below).

Particular simple cases have 
already been studied: considering a 2-dimensional \cite{D2a, D2b,
D2c} or a 3-dimensional \cite{Pozo-Escola3D} space-time, or for
special motions of the emitters in the Schwarzschild geometry from
analytical \cite{BiniMas} and numerical \cite{DelvaMas}
approaches. For a recent approach to emission coordinates 
using the integration of the eikonal equation, 
and some numerical simulations, see also Ref. \cite{Bu-Ca-Matzner-2011}.

In this and the following sections, we are mainly dealing with relativistic positioning systems in Minkowski
space-time.

\subsection{Covariant expression of the solution}
\label{subsec-2a}

\begin{figure}[b]
\centerline{
\parbox[c]{0.5\textwidth}{\includegraphics[width=0.4\textwidth]{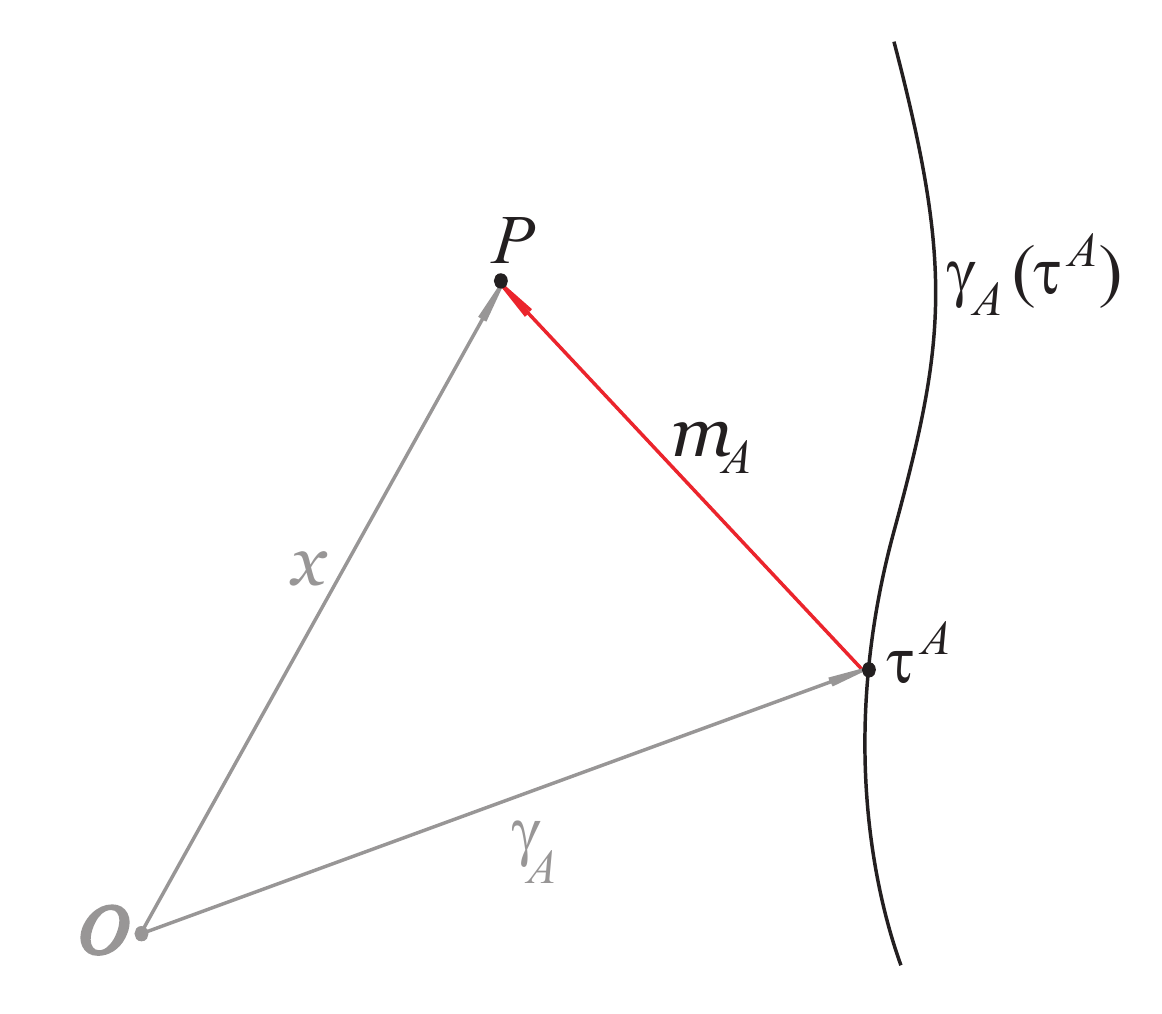}}}
\caption{The emission of an electromagnetic signal from a satellite $\gamma_A(\tau^A)$ at  proper time $\tau^A$, 
and its reception by a user at $x\equiv OP$. These events define the future pointing null vector $m_A$.
 \label{notation}}
\end{figure}

From now on, we shall suppose that any user in the emission coordinate region $\cal{C}$ receives the standard emission data set $E$.
Let us denote by $x$ the position vector (with respect the origin $O$ of this inertial system) of an event $P$ in the 
emission region $\cal{R}$, $x \equiv OP$. If a user  at $P$ receives the broadcast times $\{\tau^A\}$, $\gamma_A$ denote the 
position vectors of the emitters at the emission times,   
$\gamma_A \equiv O\gamma_A (\tau^A)$. Then
\begin{equation}\label{defellnull}
m_A \equiv x - \gamma_A, \quad \quad  (A=1, ..., 4),
\end{equation}
are future-oriented light-like vectors that represent the
trajectories followed by the electromagnetic signals from the
emitters $\gamma_A(\tau^A)$ to the reception event $x\in{\cal R}$  (see Fig. \ref{notation}).

In the standard emission data set $E$, the emission data $\{\tau^A\}$ received at $x$ are the emission
coordinates of the event $x \in {\cal R}$ and were broadcast when
the emitters were  at the events $\{\gamma_A(\tau^A)\}$, the
configuration of the emitters for the event $x$. Generically,
these four events determine the {\it configuration
hyperplane} for $x$.%
\footnote{Here, we always consider that the emitter configuration is {\it regular}, i. e., the four events
$\{\gamma_A(\tau^A)\}$ are noncoplanar.  Nonregular or {\it degenerate} configurations are considered elsewhere 
(for some remarks concerning these situations,
see Refs. \cite{Abel-Chaffee-91, Chaffee-Abel-94}).
\label{degenerate configurations}}

For the events $x$ in the emission coordinate region ${\cal C}$, the transformation $x = \kappa(\tau^A)$ from emission to
inertial coordinates is locally well defined.   In
\cite{emission-1}, we have obtained a covariant expression of this transformation,  
given by the following formula:
\begin{equation}\label{generaltransf}
x = \gamma_4 + y_* -  \frac{y_*^2 \, \chi }{(y_*\cdot \chi ) + {\hat \epsilon}\sqrt{(y_*\cdot \chi )^2 -
y_*^2 \chi^2}}
\end{equation}
where $\gamma_4(\tau^4)$ has been chosen as the reference emitter.%
\footnote{The transformation (\ref{generaltransf}) from emission to inertial coordinates may be written in a totally symmetric form without the 
choice of any emitter worldline as reference origin line. For this purpose,  one has to consider the barycenter of the emitters as 
the convenient reference event rather than one of the emitters. This issue will be addressed elsewhere \cite{Minko-baricentric}, in 
connection with the symmetric formulation of the location problem in flat space-time. 
\label{emico}}

\begin{figure}[t]
\centerline{
\parbox[c]{0.5\textwidth}{\includegraphics[width=0.4\textwidth]{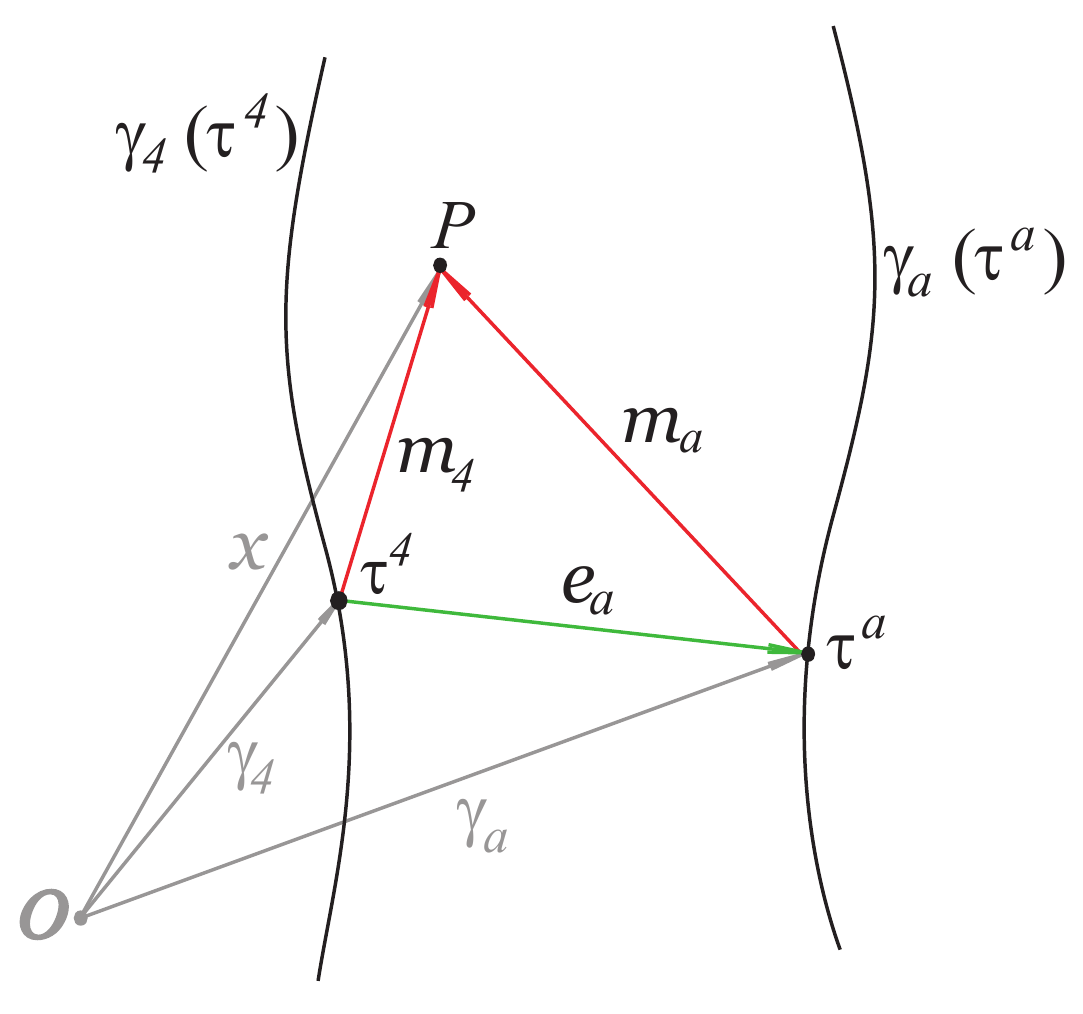}}}
\caption{If we choose the emitter four as origin (reference emitter), the relative positions of the others (referred
emitters) are $e_a = \gamma_a - \gamma_4$, $a=1,2,3$, and the
position vector of the event $P$ is $m_4$. \label{notation-2}}
\end{figure}

Quantity $y_*$ is given by
\begin{equation}\label{y-estrella-H}
y_* = \frac{1}{\xi \cdot \chi} \, i(\xi) H, 
\end{equation}
where $\chi$ is the {\it configuration vector} 
\begin{equation}\label{chi}
\chi = * (e_1 \wedge e_2 \wedge e_3)
\end{equation}
and $H$ is the {\it configuration bivector}
\begin{equation}\label{H}
H = * (\Omega_1 \, e_2 \wedge e_3 + \Omega_2 \, e_3 \wedge e_1
 + \Omega_3 \, e_1 \wedge e_2)
\end{equation}
with (see Fig. \ref{notation-2})
\begin{equation}\label{Omegas-ea-chi}
\Omega_{a} = \frac{1}{2}(e_a)^2,  \quad    e_a =
\gamma_a - \gamma_4,  \quad  (a = 1, 2, 3), 
\end{equation}
and where $\xi$ is any vector transversal to the configuration,
$\xi\cdot\chi \neq 0$, and  $i(\xi)H$ stands 
for the tensor contraction of $\xi$ and  the first slot of $H$ (see Appendix \ref{apendix}). 

Quantity ${\hat \epsilon}$  is the {\it orientation} of the
positioning system at $x$, that  is now equivalently expressed as 
\begin{equation} \label{deforientation}
{\hat \epsilon} \equiv \textit{sgn}  [* (m_1 \wedge m_2 \wedge m_3
\wedge m_4)].
\end{equation}

It is worthy to remark that $\chi$ and $H$ are determined by the relative positions $e_a = \gamma_a - \gamma_4$ 
associated with a given configuration of the emitters. Therefore, $y_*$ is directly computable from the sole standard emission data.

Nevertheless,  if we want to obtain $x$ from (\ref{generaltransf}) we  also need to
determine the orientation ${\hat \epsilon}$, which involves, by
substituting (\ref{defellnull}) in (\ref{deforientation}), the
unknown $x$. In fact, from Eqs. (\ref{defellnull}) and (\ref{Omegas-ea-chi}) it is clear that $m_a = m_4 - e_a$ (see Fig. (\ref{notation-2})) and one obtains:
\begin{equation} \label{emes-es}
m_1 \wedge m_2 \wedge m_3 \wedge m_4 = - (e_1 \wedge e_2 \wedge e_3 \wedge m_4)
\end{equation}
that taking into account (\ref{chi}) allows us to express Eq. (\ref{deforientation}) as 
\begin{equation} \label{orient-m4-chi}
{\hat \epsilon} = \textit{sgn}  \, (\chi \cdot m_4)
\end{equation}
which by (\ref{defellnull}) depends on $x$. Therefore, in order to show that Eq.
(\ref{generaltransf}) does not chase its own tail, we must be
able to determine the orientation ${\hat \epsilon}$ at $x$ by using
a procedure not involving the previous knowledge of $x$.

\subsection{Analysis of the solution}
\label{subsec-2c}

%
\begin{figure}[b]
\centerline{
\parbox[c]{0.5\textwidth}{\includegraphics[width=0.5\textwidth]{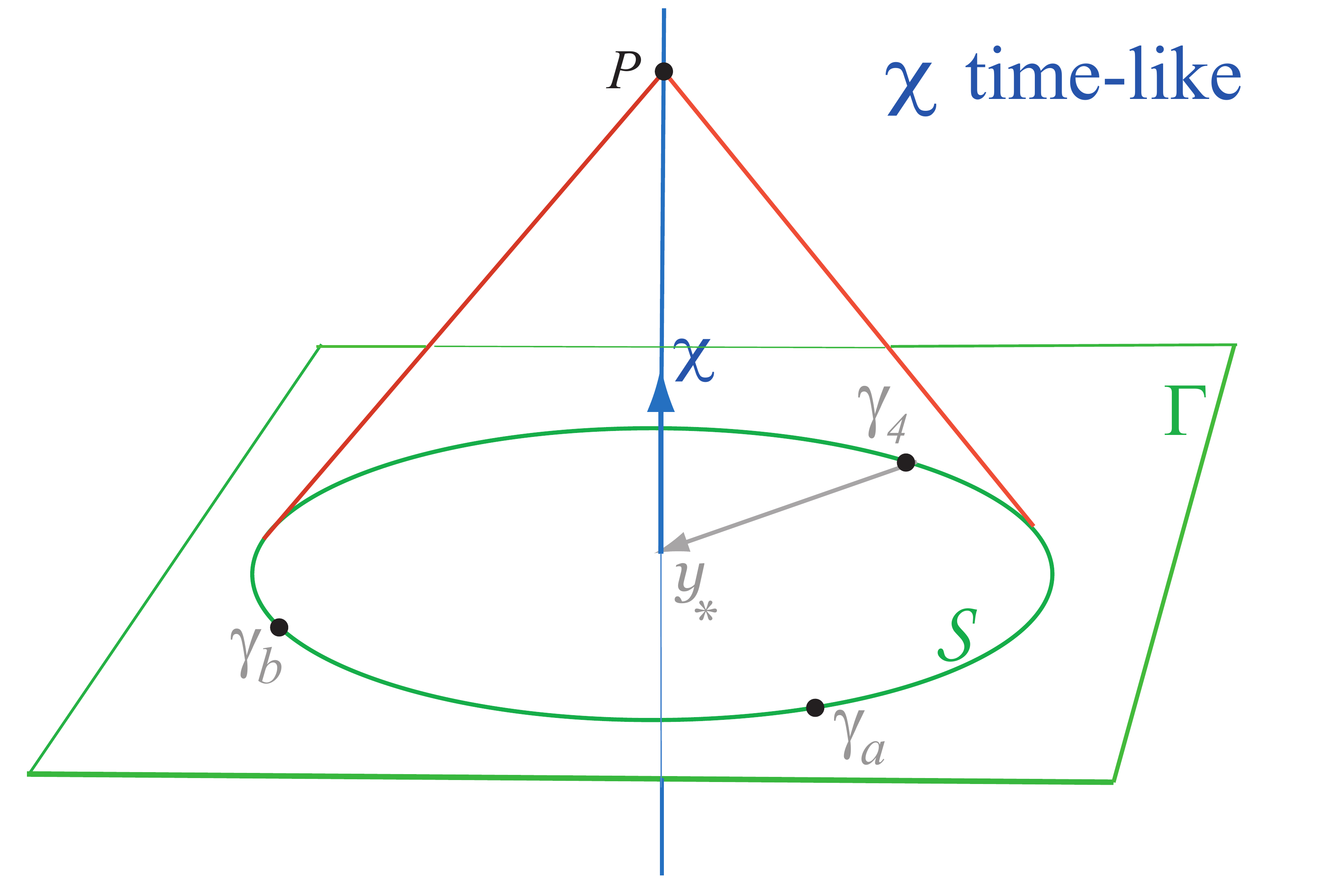}}}
\caption{For the event $P$, the configuration hyperplane $\Gamma$ is space-like, $\chi^2<0$. In this case, 
the emitters remain on a 2-sphere ${\cal S}$ laying in $\Gamma$, and a sole emission 
solution $P$ exists.  In this 3-dimensional representation for three satellites, the 2-sphere reduces to a circle.
 \label{confi-espacial}}
\end{figure}
\begin{figure}
\centerline{
\parbox[c]{0.5\textwidth}{\includegraphics[width=0.5\textwidth]{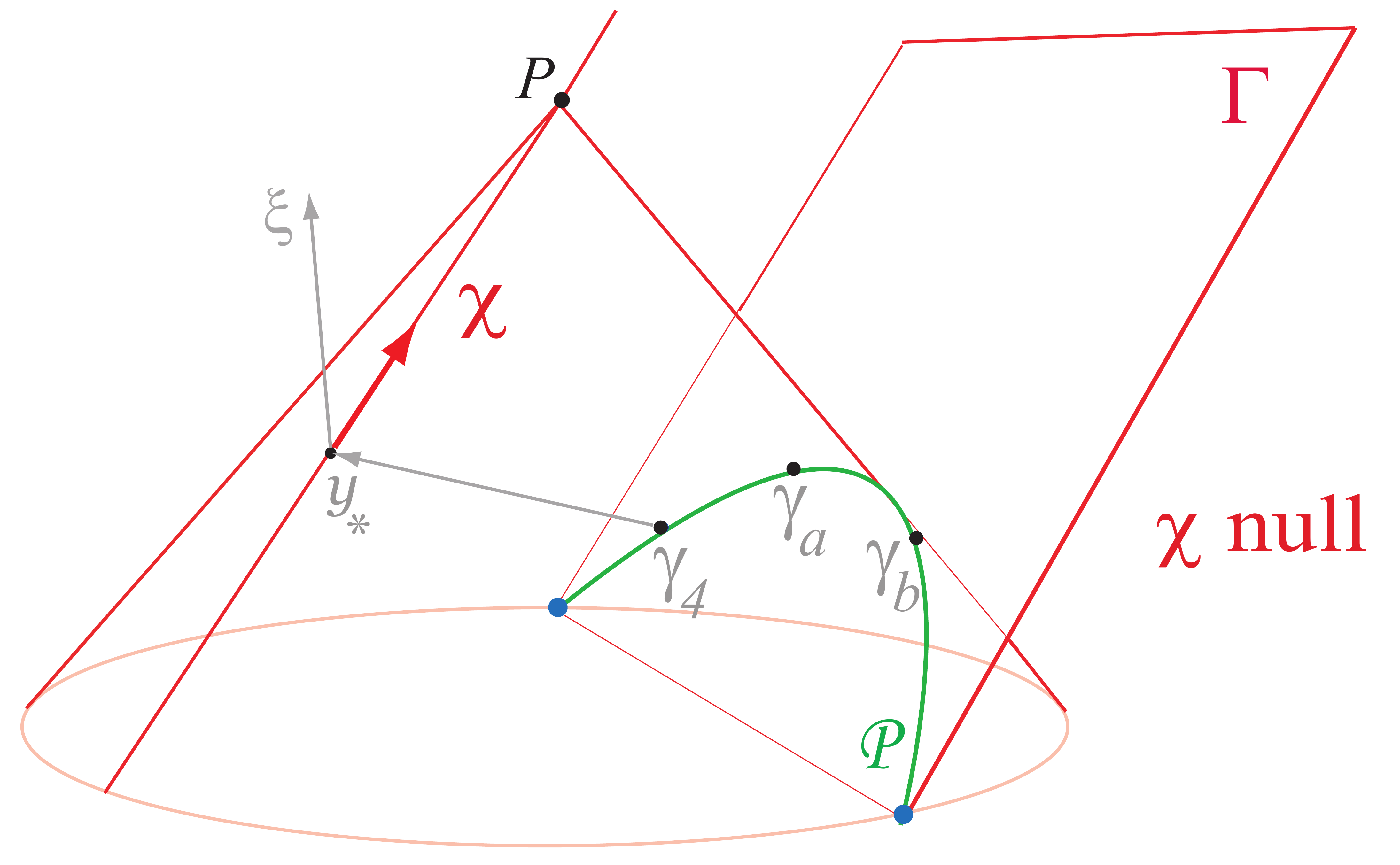}}}
\caption{For the event $P$, the configuration hyperplane $\Gamma$ is light-like,  $\chi^2=0$. In this case, 
the emitters remain on a 2-paraboloid ${\cal P}$ laying in $\Gamma$, and a sole emission 
solution $P$ exists.  In this 3-dimensional representation for three satellites, the 2-paraboloid  reduces to a parabola.
\label{confi-isotropa}}
\end{figure}

In Ref. \cite{emission-1}, Eq. (\ref{generaltransf}) was obtained by separately analyzing three different cases, and gluing together their different solutions 
in a sole covariant an analytic expression.  In gluing them,  the role played by the external element $\xi$ is essential.%
\footnote{For a detailed discussion about this point, see Ref. \cite{emission-1}.}%

The three cases correspond to the different causal characters  of the configuration vector $\chi$.  In space--time metric signature $(-, +, +, +)$, one has for each case:

(i) $\chi$  time-like,  $\chi^2  <  0$ :  there is a sole emission solution $x$ (the other one is a reception solution).  The orientation 
${\hat \epsilon}$ corresponding to the emission solution remains to be calculated; 

(ii) $\chi$  light-like,  $\chi^2 = 0$ : there is a sole emission solution $x$ (the other one being degenerate).  The orientation 
${\hat \epsilon}$ corresponding to the emission solution remains to be calculated; 

(iii) $\chi$  space-like,  $\chi^2 > 0$ :   there are two emission solutions,  $x$ and $x'$. They only differ by their 
orientation ${\hat \epsilon}$. The problem is how to determine the one corresponding to the real user.

\begin{figure}
\centerline{
\parbox[c]{0.5\textwidth}{\includegraphics[width=0.5\textwidth]{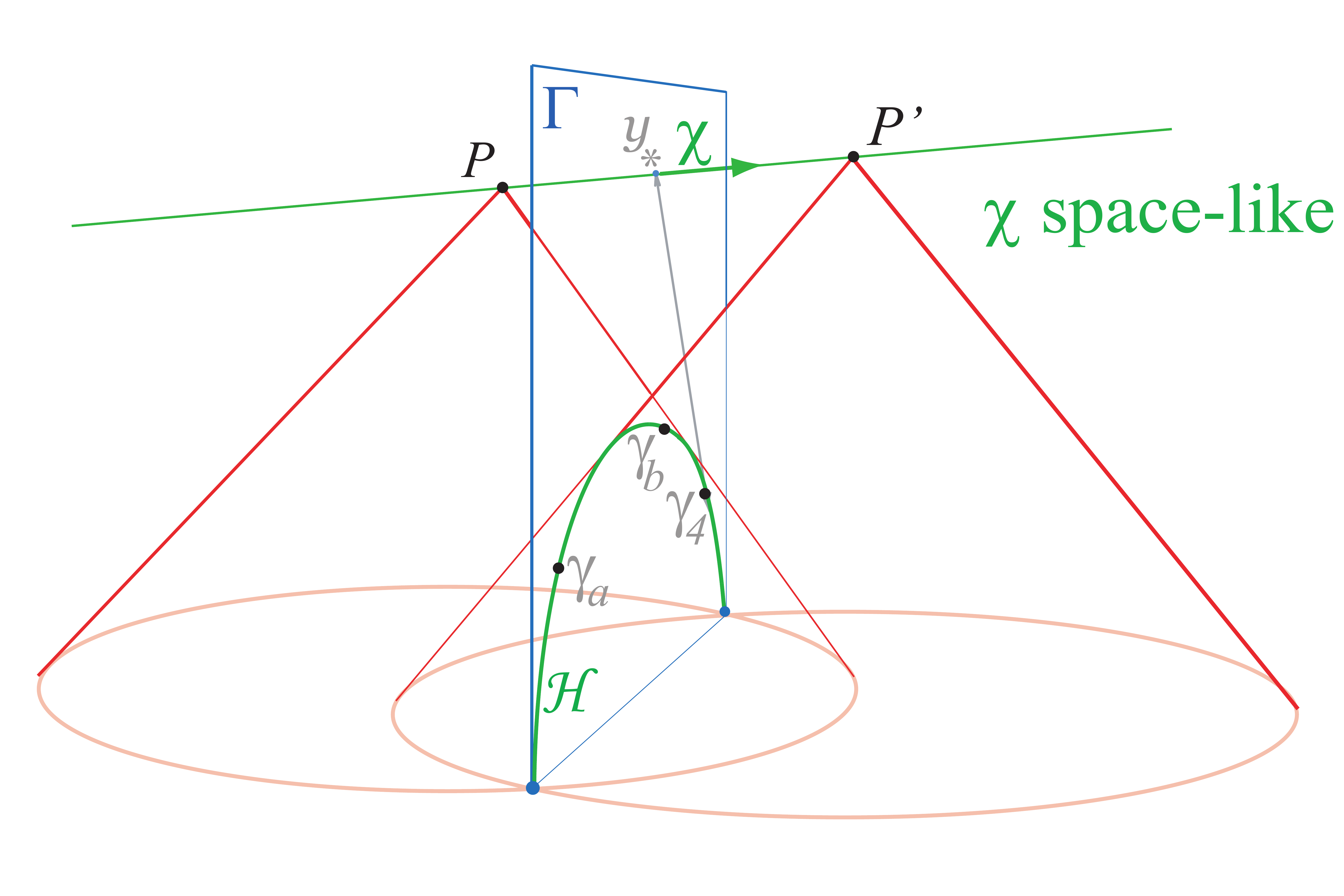}}}
\caption{For the events, $P$ and $P'$,  the configuration hyperplane $\Gamma$ is time-like,  $\chi^2>0$. In this case, 
the emitters remain on a 2-hyperboloid ${\cal H}$ laying in $\Gamma$, and both $P$ and $P'$ are emission 
solutions corresponding to the same emission coordinates ($\tau^A$).  In this 3-dimensional representation for three satellites, the 2-hyperboloid  reduces to a hyperbola.
\label{confi-temporal}}
\end{figure}

The above cases are illustrated in Figs. \ref{confi-espacial}, \ref{confi-isotropa} and \ref{confi-temporal}.

For cases (i) and (ii), the matter to determine ${\hat \epsilon}$  was solved in \cite{emission-1} (see Sec. \ref{subsec-4b} below). 
Figure \ref{confi-espacial} shows the emission solution for the case (i).  The configuration hyperplane, being space-like, cuts the past light 
cone of the solution in a 2-sphere containing the configuration of the emitters. Figure \ref{confi-isotropa} shows case (ii), where the 
emitter configuration stays on a 2-paraboloid contained in the null configuration hyperplane.

For case (iii),  ${\hat \epsilon}$ can not be determined from the sole emission data. Figure \ref{confi-temporal} shows a pair of emission solutions receiving the same emission coordinates. This `indetermination' is known as the {\it bifurcation problem}. To solve it is the main subject of this paper (see Sec. \ref{sec-5}).


\section{The border between the emission coordinate domains}
\label{sec-3}

The emission coordinate region contains two emission coordinate domains (see Sec. \ref{sec-4} below). The border between these domains is the hypersurface 
${\cal J}$, where the Jacobian determinant of the characteristic emission function $\Theta$ vanishes, 
\begin{equation} \label{J}
{\cal J} = \{x \ | \ j_\Theta(x) = 0\}.
\end{equation}
We are going to obtain some related properties showing its interest in relativistic positioning. 

First, let us note that, in an adequate and condensed form, Eq. (\ref{generaltransf}) reads as
\begin{equation}\label{condensed}
x = \gamma_4 + y_* - \lambda \chi,\end{equation}
where 
\begin{equation}\label{lambda-Delta}
\lambda \equiv \frac{y_*^2}{(y_*\cdot \chi ) + {\hat \epsilon}\sqrt{
\Delta}}, \quad \quad \Delta \equiv (y_* \cdot \chi )^2 -
y_*^2 \chi^2.
\end{equation}
As $m_4 = y_*  - \lambda \chi$ is a null vector,   and vectors  $\{y_*, \chi\}$ and   $\{m_4, \chi\}$ generate  the same 2-plane,  the following relation holds:
\begin{equation} \label{Delta-chi-m4}
sgn \, (\Delta) = sgn \,  [(\chi \cdot m_4)^2] 
\end{equation}
and then $\Delta \geqslant 0$, assuring consistence for  the above definition of $\lambda$. Consequently, one has the following result, made already evident by Eqs. (\ref{Delta-chi-m4}) and (\ref{orient-m4-chi}).

{\it Proposition 1:}  $ j_\Theta(x) = 0$  if, and only if,  $\Delta = 0$. \\

The fact that $\Delta$ is non-negative says that  the 2-plane generated by $y_*$ and $\chi$ is everywhere time-like, except in the border ${\cal J}$, where this plane is light-like.

Coming back to Eq. (\ref{H}), let us note that $H$ is a simple bivector,  that is,  
\begin{equation}\label{Hsimple}
H = \chi \wedge a
\end{equation}
for some vector $a$, because of $i(\chi) * H = 0$, which is a direct consequence of Eqs. (\ref{chi}) and  (\ref{H}).  Therefore,  
the invariant $(H, *H)$ vanishes, $(H, *H) = 0$, and the invariant $(H,H)$ takes the expression (see Appendix \ref{apendix}):
\begin{equation}\label{HH-invariant}
(H,H) = \chi^2 a^2 - (\chi \cdot a)^2.
\end{equation}

On the other hand, substituting (\ref{Hsimple}) into (\ref{y-estrella-H}), $y_*$ is expressed as%
\begin{equation}\label{y-estrella-xia}
y_* = \frac{1}{\xi \cdot \chi} \, [(\xi \cdot \chi) a - (\xi \cdot a) \chi], 
\end{equation}
and then, Eq. (\ref{lambda-Delta}) for $\Delta$ becomes
\begin{equation}\label{Delta-chi-a}
\Delta = (\chi \cdot a)^2 -  \chi^2 a^2.
\end{equation}
Consequently, $\Delta$ really does not depend on the choice of the transversal vector $\xi$ and, by comparing (\ref{HH-invariant}) and (\ref{Delta-chi-a}),  
the following result has been proved.

{\it Proposition 2:} Up to sign, the quantity $\Delta$ defined in (\ref{lambda-Delta})  
is the scalar invariant $(H, H)$ of the configuration bivector $H$:
\begin{equation}\label{Delta-invariantH}
\Delta = - (H,H).
\end{equation}

Moreover,  from Eq. (\ref{Delta-invariantH}),  the user  can determine $\Delta$ from the sole standard emission data $E$. Thus, taking into account Proposition 1, the user is able to know, from the sole standard set $E$ it receives,  when it is crossing the border  ${\cal J}$ of the two emission coordinate domains. 

Furthermore, it is worth remarking that on the border $\cal{J}$  the location of a user may be unambiguously solved. There, its location is 
obtained from (\ref{condensed}) by taking $\Delta =0$ in Eq. (\ref{lambda-Delta}).

On the way, taking into account that $(H, H) = 0$, $H$ will be a null bivector only when the invariant $(H, *H)$ vanishes. Then,  we have also proven the following result.

{\it Proposition 3:}  For an event $x \in {\cal{R}}$, the configuration bivector $H$ is a null bivector if, and only if $ j_\Theta(x) = 0$.\\

On the other hand, an {\it observational method} allowing the user to detect when it is on the border ${\cal J}$ has been previously studied by Coll and Pozo,   
who stated the following result \cite{Pozo-Escola4D,Pozo-Varsovia-2005}.

{\it Proposition 4:} The border ${\cal J}$ consists in those events for which any user at them can see the four emitters  on a
circle on its celestial sphere.\\

This result is rather counterintuitive. When the GPS satellites are all  near the horizon, or are all too close together 
on our zenith, the error in positioning is great. It would seem then that  the optimal conditions for a precise location  
would be obtained when all the satellites are situated on an intermediate circle of the celestial sphere (say, among 30 or 60 degrees with respect to the zenith).  Nevertheless, Proposition 4 shows that the circle corresponds to the most degenerate distribution that a set of satellites may have.

Proposition 4 also makes clear that the border ${\cal J}$ may be plotted 
from the sole observational data, a result that was not, {\it a priori}, evident.


\section{Regions and coordinate domains in relativistic positioning}
\label{sec-4}

This section provides a geometrical background to analyze the space-time regions which are relevant in 
relativistic positioning. In particular, we study the subset of the emission coordinate region ${\cal C}$, 
 where the orientation ${\hat \epsilon}$ is computable from the standard emission data $E$
 (the central region of the positioning system).

\subsection{Emission configuration regions $\mathcal{C}_s, \mathcal{C}_\ell$, and $\mathcal{C}_t$}
\label{subsec-4a}

The emission coordinate region  ${\mathcal{C}}$ is constituted by three disjoint
regions, and one can write $\mathcal{C} = \mathcal{C}_s \cup \mathcal{C}_\ell \cup \mathcal{C} _t$.
They are the {\it space-like} $\mathcal{C}_s$, the {\it null} $\mathcal{C}_\ell$ and the {\it time-like} $\mathcal{C} _t$ emission 
configuration regions defined by the conditions $\chi^2 < 0$, $\chi^2 = 0$, and $\chi^2 > 0$, respectively.

This means that at every event $x \in \mathcal{C}_s$ ($x \in \mathcal{C}_\ell$ or  $x \in \mathcal{C}_t$, respectively) 
a user receives the signals from four emission events that generate a space-like (null or time-like, respectively) hyperplane.

From Eqs. (\ref{chi}) and (\ref{Omegas-ea-chi}),  which only involve the emitter configuration of the standard emission data $E$, 
the user is able to determine the sign of $\chi^2$. Consequently, from the data set $E$ and the above definitions, 
the user knows in what configuration region, 
${\cal C}_s$, ${\cal C}_\ell$, or ${\cal C}_t$, of the positioning system it is traveling.

\subsection{The central region ${\cal C}^C = {\cal C}_s \cup {\cal C}_\ell$}
\label{subsec-4b}

We name ${\cal C}^C \equiv {\cal C}_s \cup {\cal C}_\ell$ the {\it central region} of the positioning system. 

At every event $x \in {\cal C}^C$, one has $u\cdot \chi \neq 0$ for any future pointing time-like vector $u$,
because  $\chi$ is not space-like in this region. Taking into account that $m_4$ is a future pointing null vector,   
the sign of the scalar products  $\chi \cdot m_4$ and $u\cdot \chi$ is the same for any  future pointing time-like vector $u$,  
and from Eq. (\ref{orient-m4-chi}) this sign is precisely the orientation of the positioning system on the central region. 
More precisely,  we can prove the following result:

{\it Proposition 5:} In the central region  ${\cal C}^C$, the orientation of a relativistic positioning system 
is constant, and may be evaluated from the sole standard emission data $E$:
\begin{equation} \label{orientacio-central}
\forall x \in {\cal C}^C,  \qquad {\hat \epsilon} = \textit{sgn}\,(u\cdot\chi)
\end{equation}
where $u$ is any future pointing time-like vector. \\

Thus, from (\ref{orientacio-central}) any user in the central region is
able to determine the orientation of the positioning system, and then, from Eqs. (\ref{generaltransf})-(\ref{Omegas-ea-chi}),  it  
can obtain its own position $x$ in the inertial system from the sole standard emission data by
substituting ${\hat \epsilon} = \textit{sgn}\,(u\cdot\chi)$ in
(\ref{generaltransf}). The resulting sign of ${\hat \epsilon}$ will be positive or negative depending on the time orientation 
of the computed vector $\chi$.

\subsection{Front (${\cal C}^F$) and back (${\cal C}^B$) coordinate domains}
\label{subsec-4c}

As a consequence of  Proposition 5,  the Jacobian determinant does not vanish, $ j_\Theta(x) \neq 0$, in the immediate vicinity of 
$\cal{C}^C$. Therefore, the border ${\cal J}$ divides  the time-like configuration region ${\cal C}_t$  of ${\cal C}$. In other words, the whole region  ${\cal C}_t$ cannot be recovered by a sole coordinate domain.%
\footnote{Remember that, for historical reasons, the {\em coordinate domain} of a coordinate system is an {\em open } set, but not necessarily 
a {\em topological domain} (i. e. an open and connected set).}  In \cite{emission-1} we proved the following result.

{\it Proposition 6:}  The emission coordinate region ${\cal C}$ is not a
coordinate domain but the union of two disjoint coordinate
domains, called the front ${\cal C}^F$,
and the back ${\cal C}^B$ emission coordinate domains, ${\cal C} = {\cal C}^F \cup {\cal C}^B$. \\

The front coordinate domain ${\cal C}^F$ contains the central region ${\cal C}^C$ and    
a proper subset  ${\cal C}_t^F$ of the time-like configuration region, ${\cal C}_t^F \subset {\cal C}_t$. 
This proper subset ${\cal C}^F_t$  is the part of ${\cal C}_t$ adjacent to the central region ${\cal C}^C$, so that the whole front domain 
${\cal C}^F$, ${\cal C}^F \equiv {\cal C}^C \cup  {\cal C}^F_t$, has, by continuity, constant orientation ${\hat \epsilon}$ 
(the same as the central region). However,  the orientation at $x \in {\cal C}_t^F$ can not be determined from Eq. (\ref{orientacio-central}), 
because Proposition 5 only applies on ${\cal C}^C$.

The back coordinate domain ${\cal C}^B$ is not a simply-connected domain. In fact, the region ${\cal C}_t$ is not simple-connected, and its leaves are constituted by pairs of events $\{x, x'\}$ having the same emission coordinates but different inertial ones, defining two well 
differentiated regions:  if $x \in {\cal C}^B$, then $x' \in {\cal C}_t^F \equiv  {\cal C}_t - {\cal C}^B$.

\begin{figure}
\centerline{
\parbox[c]{0.5\textwidth}{\includegraphics[width=0.4\textwidth]{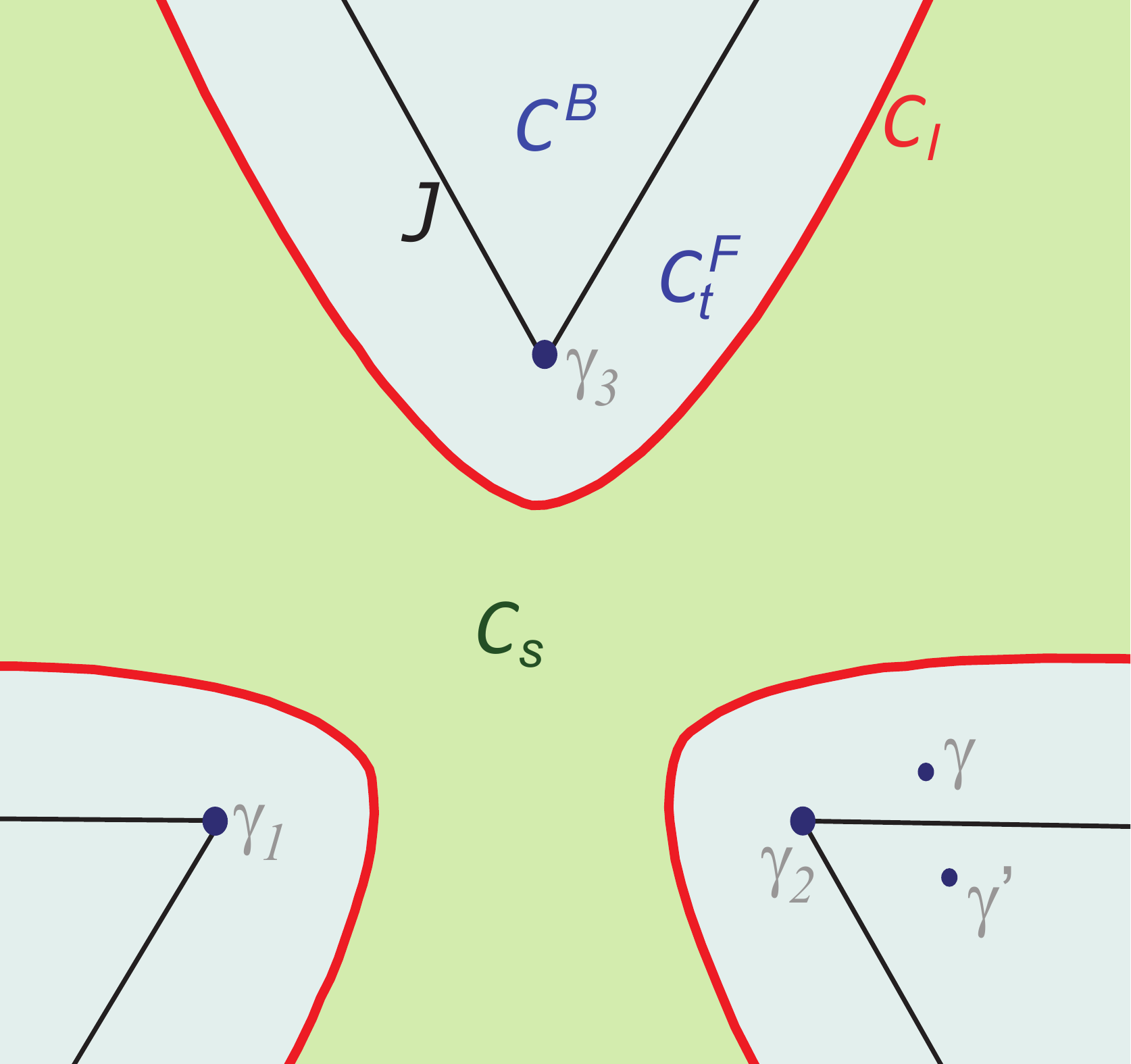}}}
\caption{Quotient space $S$ of the stationary observers of a stationary and symmetric relativistic positioning system in flat 3-dimensional space-time.  
Dots $\gamma, \gamma' \in S$ represent stationary user worldlines, and solid dots $\gamma_a$ stand for stationary satellites (emitters).
The emission configuration regions ${\cal C}_s$, ${\cal C}_\ell$ and ${\cal C}_t$ are differently colored. 
The border between ${\cal C}^B$ and ${\cal C}^F_t$ is ${\cal J}$: the surface of vanishing Jacobian.  
Conjugate events, $(x, x')$, having the same emission coordinates but different inertial ones, necessarily occur on
separate parts of ${\cal C}_t$: if $x \in \gamma \subset  {\cal C}_t^F$, then $x' \in \gamma' \subset {\cal C}^B$. 
In this 3-dimensional situation, ${\cal J}$ is just the union of the shadows that each satellite produces on the signals coming 
from the other ones in the region  ${\cal C}$. \label{domains}}
\end{figure}

To illustrate these coordinate domains, let us consider the simple case of a symmetric stationary positioning system in flat space-time. 
In this case, the four emitters define a regular tetrahedron,  and ${\cal C}^B$ is
the union of four connected components. The common boundary ${\cal J}$ of the domains ${\cal C}^F$ and ${\cal C}^B$ is a 
four-leaf hypersurface that contains the shadows that each satellite produces on the signals coming from the other ones in the 
region  ${\cal C}$. The orientation ${\hat \epsilon}$ of the positioning system only changes 
across ${\cal J}$ taking different constant value on each coordinate domain. The analogous, but simpler to draw, 
stationary and symmetric 3-dimensional case is illustrated in Fig. \ref{domains}, that shows the involved configuration regions and coordinate domains.

%
%


%
\section{The bifurcation problem. Its observational solution}
\label{sec-5}
\begin{figure*}
\centerline{
\parbox[c]{0.75\textwidth}{\includegraphics[width=0.8\textwidth]{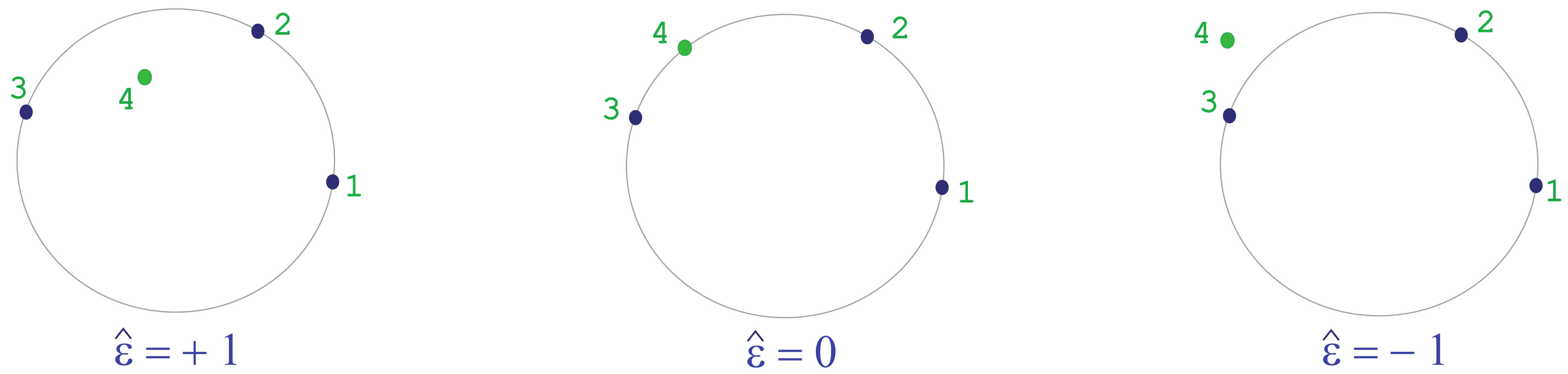}}}
\caption{Observational rule to determine the orientation ${\hat \epsilon}$ at the location of a user from the relative positions of the emitters on its celestial sphere.  In the left diagram, the visual axis is oriented towards the spherical cap that does not contain the fourth emitter, and then  ${\hat \epsilon} = + 1$.  In the middle diagram, the four emitters are on a circle of the celestial sphere of the user, and thus ${\hat \epsilon} = 0$ (Proposition 4). In the right diagram, the visual axis is oriented towards the spherical cap that contains the fourth emitter, and thus ${\hat \epsilon} = - 1$.  \label{orientacio}}
\end{figure*}

The above results show that the standard emission data $E$ are generically insufficient to locate a user 
of a positioning system in an inertial system. 

In the past, and in connection with GNSS, this problem was pointed out by Schmidt  \cite{Schmidt-72} and studied 
by Abel and Chaffee \cite{Abel-Chaffee-91,Chaffee-Abel-94} by introducing a {\rm ``bifurcation parameter"} (equivalent to the square $\chi^2$ of the configuration vector $\chi$ of Eq.  (\ref{chi})). Afterwards, it was referred as the {\em bifurcation problem} \cite{Grafarend-Shan-1996}.

In current practical situations in present day GNSS,%
\footnote{A present day GNSS allows locating only part of the interior of a sphere surrounding the satellite constellation.}%
\, the bifurcation problem may be solved by hand: simply checking which of the two solutions satisfies an observable pertinent constraint. 
Thus, for example, if a user stays near the Earth surface the right solution is the nearest to the Earth radius. However, in extended GNSS or more general positioning systems in the Solar System, the bifurcation problem cannot be so easily avoided; it will always be present for users traveling in the time-like configuration region ${\cal C}_t$.

One could think that the bifurcation problem could be avoided by continuity for users traveling from the central region  ${\cal C}^C$
(where they are able to calculate the positioning system orientation from the standard emission data) to the time-like configuration region ${\cal C}_t$. 
But the discrete character of true successive location operations, and the fact that it suffices of a sole instant to  cross the border ${\cal J}$, 
also make this possibility illusory. Not only for theoretical reasons, but also for future practical applications where the 
role played by Earth based coordinate systems could become secondary (cf. \cite{Coll-ERE-2000,D4a,D2a}), it is 
essential to learn to solve this important part of the location problem, the bifurcation problem. 

We have seen that, from the sole standard emission data $E$, the users can know the configuration region that they are traveling. 
The bifurcation problem appears when this configuration region is the time-like one, ${\cal C}_t$, because this region is constituted 
by pairs of conjugate events, $x$ and $x'$, separated by the border 
${\cal J}$, receiving the same standard emission data (see Figs. \ref{confi-temporal} and \ref{domains}).  
Conjugate events belong to different (back and front) coordinate domains, of different orientation. As Eq. (\ref{generaltransf}) shows, 
the knowledge of the  orientation in addition to the data set $E$ solves completely the bifurcation problem. 
Thus, how to extend the standard emission data $E$ so as to be able to determine the orientation of the positioning system for the user?

We shall suppose here that, in addition to the standard emission data $E$, 
the users are able to {\em observe} the relative positions of the emitters on their celestial sphere. 
We shall denote this {\em extended data set} by $E^*$.

Consider an arbitrary user of unit velocity $u$,  at the event $x$
of ${\cal C}$.  With respect to this user, the null vectors $m_A$ may be decomposed as
\begin{equation}
\label{eleAelebarraA}  m_A =  (m_A)_u u + \vec{m}_A, 
\end{equation}
where $(m_A)_u = - u\cdot  m_A > 0$ and $\vec{m}_A $ denote vectors of the proper 3-dimensional space 
$S_u$ of the user, $\vec{m}_A \in S_u$ (cf. Eq. (\ref{descompo})).

Let us consider the unit vectors $\vec{n}_A = \vec{m}_A / (m_A)_u$ 
giving the relative directions of propagation of the signals.
Because the vectors $- \vec{n}_A$ point to the positions of the  emitters $A$, 
i.e.,  are the unit vectors along the apparent line of sight of the emitters $A$,
we say that  $\{\vec{n}_A\}$ is a set of {\it observational data}. 
It is this set of data (or any equivalent one) which, added to the standard emission data, is included in  $E^*$.

By direct substitution of (\ref{eleAelebarraA}) in the expression
(\ref{deforientation}) of ${\hat \epsilon}$, one has 
\begin{eqnarray} \label{signJacob} 
{\hat \epsilon} =  \textit{sgn} \, \{* && [u \wedge ( -\vec{n}_1 \wedge \vec{n}_2 \wedge \vec{n}_3  + 
\vec{n}_1 \wedge  \vec{n}_2 \wedge \vec{n}_4 \nonumber \\
&& - \vec{n}_1 \wedge \vec{n}_3 \wedge \vec{n}_4 + \vec{n}_2 \wedge \vec{n}_3 \wedge  \vec{n}_4)]\}
\end{eqnarray}
where we have taken into account that 
$\prod_{A=1}^4 m_A^0 > 0$ for emission vectors $ m_A$. And because any 3-form ${\cal F}$ in $S_u$ satisfies 
$i(u)*{\cal F}  = *(u\wedge {\cal F})$, the above expression gives (see Appendix \ref{apendix}), 
\begin{eqnarray} \label{signJacob1-1} 
{\hat \epsilon} =  \textit{sgn} \, [&& (\vec{n}_1, \vec{n}_2, \vec{n}_3) - (\vec{n}_2, \vec{n}_3, \vec{n}_4) \nonumber \\
&& + (\vec{n}_3, \vec{n}_4, \vec{n}_1) - (\vec{n}_4, \vec{n}_1, \vec{n}_2)], 
\end{eqnarray}
where the triple product is defined according to Eqs. (\ref{triple-p}) and (\ref{u-triple-p}). Then, the following result holds.
{\it Proposition 7:} The orientation ${\hat \epsilon}$ of a relativistic positioning system is given by 
\begin{eqnarray} \label{signJacob2} 
{\hat \epsilon} =  \textit{sgn} \, [(\vec{\mu}_1, \vec{\mu}_2, \vec{\mu}_3)]
\end{eqnarray}
with $\vec{\mu}_a \equiv  \vec{n}_a - \vec{n}_4$.\\

Thus, from the relative positions of the emitters on the celestial sphere of the user, we can obtain the orientation ${\hat \epsilon}$. 
 For instance, if the referred emitters 1, 2, 3 are counterclockwise aligned on a circle of the celestial sphere of the user and the fourth emitter is inside this circle, then $(\vec{\mu}_1, \vec{\mu}_2, \vec{\mu}_3) > 0$. Then, analyzing separately all the possible situations
 we arrive to the following rule to obtain the orientation.\\

{\it Observational rule to determine ${\hat \epsilon}$.}  For any user in the coordinate region ${\cal
C}$ receiving the extended data set $E^*$, the orientation ${\hat \epsilon}$ of the positioning system may be
obtained as follows:

i) consider the circle of the celestial sphere of the user
containing the three referred emitters, $a = 1, 2, 3$,

ii) turn this circle around its center in the increasing sense $1\to 2 \to3$ to orient the visual axis of the user by the rule of the right-hand screw,  

iii) if the fourth emitter $A = 4$ is in the spherical cap pointing out by this 
oriented axis, then the orientation is ${\hat \epsilon} = -1$, otherwise ${\hat \epsilon} = + 1$. \\

By applying this observational rule, the users receiving the extended emission 
data set $E^*$ can determine the orientation ${\hat \epsilon}$ and, 
from Eq. (\ref{generaltransf}), their position in inertial coordinates.

For a better geometric comprehension of the above observational rule, 
we can  consider an alternative approach to its proof. Indeed, 
let us focus on the generic situation in which  $\{\vec{n}_a\}$ is a basis of $S_u$. Then 
the solution of the linear system
 \begin{equation}
 \label{sistemeta}
 \vec{n}_a  \cdot \vec{s}= \omega_a,  \qquad  (a=1,2,3)
 \end{equation}
is given by $\vec{s}= {\omega_a \vec{L}^a}$, with the vectors $\vec{L}^a$ 
expressed in terms of the know data $\vec{n}_A$ as
 \begin{equation}
 \label{dualbarell}
 \vec{L}^a = \frac{\epsilon^{abc}  \vec{n}_b \times \vec{n}_c}
 {2 (\vec{n}_1, \vec{n}_2, \vec{n}_3)}.
 \end{equation}
Now, by substituting (\ref{dualbarell}) into (\ref{signJacob1-1})  
we arrive to the following expression for the orientation, 
 \begin{equation}
 \label{sign3}
{\hat \epsilon} = \textit{sgn} \, [ (1 -  \vec{n}_4 \cdot \vec{L}) (\vec{n}_1,  \vec{n}_2, \vec{n}_3)]
 \end{equation}
 where $\vec{L} \equiv \vec{L}^1 + \vec{L}^2 + \vec{L}^3$.

That $\vec{n}_4 \cdot  \vec{L} = 1$ when and only when the Jacobian
$j_\Theta(x)$ vanishes has been known since \cite{Pozo-Escola4D, Pozo-Varsovia-2005}.
From Eq. (\ref{sign3}), and according to 
the result stated in Proposition 4,  
the events of the emission coordinate region $\cal C$
are all those for which the four emitters are not aligned on a
circle of the celestial sphere of the users at these events. 

Then, it is possible to state that the factor $(\vec{n}_4 \cdot  \vec{L} -1)$
in (\ref{sign3}) is positive  or negative if $\vec{n}_4$ is interior
or exterior, respectively, to the oriented half cone containing the three emitters
$\{\vec{n}_a\}$ ($a=1,2,3$). The unit vector axis $\vec{s}$ of this cone is given by
\begin{equation}
\vec{n}_a \cdot \vec{s} = \cos \varphi > 0.
\end{equation}

Moreover, in terms of the basis $\{\vec{L}^a\}$ given by Eq. (\ref{dualbarell}), the unit axis $\vec{s}$ has the expression
\begin{equation}
\label{valors} \vec{s} =  \vec{L}  \cos\varphi,
\end{equation}
as can be directly verified. 

Therefore,  a unit vector $\vec{v}$ is in the interior of the half cone or at its
exterior if the quantity $\vec{v} \cdot \vec{s}$  is greater or less than $\cos \varphi$,  
respectively,  or by (\ref{valors})  
if $\vec{v}  \cdot \vec{L} > 1$, or $\vec{v}  \cdot \vec{L} < 1$. Thus, 
by taking $\vec{v} = \vec{n}_4$, from (\ref{sign3}) one has the following result. \\

{\it Proposition 8:} Consider the oriented half-cone containing $\vec{n}_1$, $\vec{n}_2$ and $\vec{n}_3$. 
If $\vec{n}_4$ is in  its interior, then ${\hat \epsilon} =  - \textit{sgn} [(\vec{n}_1, \vec{n}_2, \vec{n}_3)]$. 
Otherwise,  ${\hat \epsilon} = \textit{sgn} [(\vec{n}_1, \vec{n}_2, \vec{n}_3)]$.\\

From this proposition we can recover the observational rule by considering all the possible relative positions of the unit vectors 
$\vec{n}_1, \vec{n}_2, \vec{n}_3$. Fig. \ref{orientacio} illustrates the application of the rule when $\{\vec{n}_1, \vec{n}_2, \vec{n}_3\}$  
is a negative-oriented basis of $S_u$, that is for $(\vec{n}_1, \vec{n}_2, \vec{n}_3) < 0$.

Let us remark that the relative positions of the emitters in the celestial sphere of a user 
are {\em Lorentz invariant}: by Lorentz transformations between users at an event, the diameter of the circle  
as well as the positions of the emitters on it  may change, but their increasing sense as well as the interior 
or exterior position of the fourth emitter will remain unchanged. 
%


\section{Discussion and ending comments}
\label{sec-concluding}

The main result of this paper is the observational rule giving the orientation ${\hat \epsilon}$ of the emission coordinates for the user. 
Together with the standard emission data, it gives a full operational character to formula (\ref{generaltransf}), allowing any user 
to obtain the coordinate transformation from emission coordinates to inertial ones and, in particular, to locate itself in inertial coordinates. 

In the central region ${\cal C}^C$, where the orientation may be {\em deduced} from the sole standard emission data (Proposition 6),  both the observed and the computed orientations may be contrasted.

It is worth to remark here that the sole standard emission data allows the users to detect when they are on the border $\cal{J}$ separating the two coordinate domains (Proposition 1), a situation that may be also contrasted with the limit of the observational rule (when the four emitters are on a circle of the celestial sphere of each user). In spite of the fact that the border does not belong to any coordinate domain, the user can also locate itself in it (taking $\Delta = 0$ in Eqs. (\ref{condensed}) and (\ref{lambda-Delta}).

Relativistic positioning concepts have been recently implemented in an algorithm giving the Schwarzschild coordinates of the users 
in terms of their emission coordinates (see \cite{DelvaMas}). If the conditions of applicability of our rule are given (observation of the emitters), 
the rule extends the region of validity of this algorithm. 

It is important to note that,  in dealing with approximate methods, or iterative algorithms, to solve the location problem in weak gravitational fields, 
Eq. (\ref{generaltransf}) is the best zero order solution to start with. 

A numerical analysis of the quantities appearing in (\ref{generaltransf}) has been recently implemented \cite{Neus-Diego-ERE10, Neus-Diego}. 
This analysis provides a numerical test of the results obtained in \cite{emission-1}, and a promising via to deal with 
numerical simulations in modeling GNSS by starting from a fully relativistic conception.


\begin{acknowledgments}
This work has been supported by the Spanish
Ministerio de Ciencia e Innovaci\'on, MICINN-FEDER Project No.
FIS2009-07705.

\end{acknowledgments}

\appendix
  
\section{Notation}
\label{apendix}

The notation of this paper has not been chosen for academic reasons but for practical ones. This notation allows, generically,  more compact and shorter expressions (in occupied space and in expended time) than index notation, improving a best understanding of the formula, but overall, it suggests more compact and shorter calculations. In this subject, where expressions and calculations are determined to become more and more complicated, the choice of appropriate symbols from the beginning is more than a matter of habit or of preference. Almost all our expressions have been calculated many times in different notations, including the index one, and the symbols in the manuscript have been chosen as the best ones from the above criteria. For readers for which this notation is not usual, we indicate here the relation between tensor index notation and ours.

(i) {\em Interior or contracted product:} $i(x)T$ denotes the contraction of a vector $x$  and the first slot of a tensor $T$. 
For instance, if $T$ is a covariant 2-tensor, $[i(x)T]_\nu = x^\mu T_{\mu\nu}$.  

(ii) {\em Exterior or wedge product:} $\wedge$. If $A$ and $B$ are both covariant (or contravariant) antisymmetric tensors, the wedge product $A \wedge B$ 
is the anti-symmetrized tensorial product $A \otimes B$. For instance,   for vectors $x$ and $y$, one has 
\begin{equation}\label{wedge1}\nonumber 
(x \wedge y)^{\mu\nu} = x^\mu y^\nu - y^\mu x^\nu, 
\end{equation}
and,  for a covector $\theta$ and a 2-form $F$, 
\begin{equation}\label{wedge2}\nonumber  
(\theta \wedge F)_{\alpha\beta\gamma} = \theta_\alpha F_{\beta\gamma} + \theta_\beta F_{\gamma \alpha} + \theta_\gamma F_{\alpha\beta}.
\end{equation}

(iii) {\em Space-time metric: $g$}, defined as a four dimensional Lorentzian metric, has components $g_{\mu\nu}$ ($\det g_{\mu\nu} < 0$). 
The signature of $g$ is taken here as $( -, +,  +, + )$. 

The scalar or inner product of two vectors $x$ and $y$ is denoted as  $x \cdot y   \equiv g(x, y) = g_{\mu\nu} x^\mu y^\nu = x^\mu y_\nu$ (in particular,  $ x^2 \equiv x \cdot x$), 
 and it is naturally extended to bivectors, $X$ and $Y$,  according to $(X, Y) \equiv \frac {1}{2} X^{\mu\nu} Y_{\mu\nu}$. Indices are raised or lowered by using the metric $g$.

(iv) {\em Metric volume element:} $ \eta$, given by
\begin{equation}\label{eta}\nonumber 
\eta_{\alpha \beta \gamma \delta} = - \sqrt{- det \, g}\, 
\epsilon_{\alpha \beta \gamma \delta},
\end{equation}
where $\epsilon_{\alpha \beta \gamma \delta}$ stands for the  Levi-Civita permutation symbol, $\epsilon_{0123} = 1$.

(v) {\em Hodge dual operator:} $*$. Let $x$ be a vector, $H$ a 2-form, and ${\cal F}$ an 3-form. 
Their associated Hodge duals, the 3-form $*x$, the 2-form $*H$, and the 1-form $*{\cal F}$,  are respectively given as
$(*x)_{\alpha\beta\gamma}= \eta_{\alpha\beta\gamma\mu} x^\mu$,  
$(* H)_{\alpha\beta} = \frac{1}{2} \eta_{\alpha\beta\mu\nu} H^{\mu\nu}$, and 
$(*{\cal F})_{\alpha} = \frac{1}{3!} \eta_{\alpha\beta\gamma\delta} {\cal F}^{\beta\gamma\delta}$.

If $x, y, z, w$ are space-time vectors, one has
\begin{equation}\label{dual1}\nonumber
[*(x \wedge y)]_{\alpha\beta} = \frac{1}{2}\eta_{\alpha\beta\mu\nu} (x
\wedge y)^{\mu\nu} = \eta_{\alpha\beta\mu\nu} x^\mu y^\nu,
\end{equation}
\begin{equation}\label{dual2}\nonumber
[*(x \wedge y \wedge z)]_\alpha = \eta_{\alpha\beta\gamma\delta} x^\beta y^\gamma z^\delta, 
\end{equation}
and
\begin{equation}\label{dual3}\nonumber
*(x \wedge y \wedge z  \wedge w) = \eta_{\alpha\beta\gamma\delta} x^\alpha y^\beta z^\gamma w^\delta.
\end{equation} 

(vi) {\em Invariants associated with a 2-form $H$}.  A space-time 2-form $H$ has associated two  independent invariants, $(H,H)$ and $(H, *H)$, which are given as: 
\begin{equation}\label{H-invariants}\nonumber
(H,H) \equiv \frac{1}{2} H^{\mu\nu} H_{\mu \nu}, \quad  (H, *H) \equiv \frac{1}{2} H^{\mu\nu} (* H)_{\mu\nu}.
\end{equation}

(vii) {\em Relative splitting}. For an arbitrary user of a relativistic positioning system (space-time observer of unit 4-velocity $u$, $u^2= -1$),  
any space-time vector $m$ may be written as: 
\begin{equation}\label{descompo}
m = m_u u +  \vec{m} 
\end{equation}
where $m_u =  - m \cdot u$  and  $\vec{m} \in S_u$ 
are the time-like and space-like components, respectively, of $m$ relative to $u$, $\vec{m} \cdot u = 0$.

(viii) {\em Induced volume on $S_u$}. 
The 3-dimensional Euclidean space orthogonal to $u$, $S_u$, has 
induced volume element, $\eta_u$, given by $\eta_u \equiv -i(u)\eta$, that is,  
$(\eta_u)_{\beta\gamma\delta} = -u^\alpha \eta_{\alpha\beta\gamma\delta}$.
The Hodge dual operator with respect $\eta_u$ is denoted as $*_u$.

(ix) {\em Cross and triple products in $S_u$}.
Vectors in $S_u$ are denoted with an arrow  above them. Thus,  
for vectors $\vec{a}, \vec{b} \in S_u$, the vector or cross product is
expressed as
\begin{equation}\label{u-dual}
\vec{a} \times\vec{b} = * (u \wedge \vec{a} \wedge \vec{b}) = *_u (\vec{a} \wedge\vec{b}).
\end{equation}
The scalar triple product  is then given by 
\begin{equation}\label{triple-p}
(\vec{a}\times\vec{b}) \cdot \vec{c} \equiv (\vec{a}, \vec{b}, \vec{c}) =  *_u (\vec{a} \wedge \vec{b} \wedge\vec{c}), 
\end{equation}
or, equivalently, 
\begin{equation}\label{u-triple-p}
 (\vec{a}, \vec{b}, \vec{c}) \, u =  * (\vec{a} \wedge \vec{b} \wedge\vec{c}).
\end{equation}

\bibliography{apssamp}

\end{document}